\documentclass[apl,twocolumn,preprintnumbers,amsmath,amssymb,a4paper]{revtex4}
\usepackage{graphicx}

\begin{document}
\title{Superconducting $\pi$ qubit with three Josephson junctions}
\author{T. Yamashita,$^{1}$, S. Takahashi,$^{1,2}$ and S. Maekawa$^{1,2}$}
\affiliation{$^{1}$Institute for Materials Research, Tohoku University, 
Sendai, Miyagi, 980-8577, Japan \\
$^{2}$CREST, Japan Science and Technology Agency (JST), Kawaguchi, Saitama, 332-0012, Japan}
\begin{abstract}
We propose a new qubit consisting of a superconducting ring with 
two ordinary zero junctions and one ferromagnetic $\pi$ junction.  
In the system, two degenerate stable states appear in the phase space without 
an external magnetic field because of a competition between the zero and $\pi$ states.  
Quantum tunneling between the two degenerate states leads to 
a formation of bonding and antibonding states (coherent states) which are used as a bit.  
For manipulating the states of the qubit, 
only small external magnetic field around zero is required.  
This feature leads to a large-scale integration and 
a construction of the qubit with a smaller size 
which is robust to the decoherence by external noises.  
\end{abstract}
\maketitle
Quantum computing has attracted a great deal of interest in the recent years \cite{Nielsen}.  
As a candidate for an elemental unit of a quantum computer (qubit), 
many proposals have been done, e.g., photons, ion traps, and nuclear spins.  
Among the proposals, solid-state devices have great advantages in 
large-scale integration and flexibility of layout, 
but still face the problem of reducing the decoherence effect 
due to their coupling to the environment.  
Recently, there are many works on superconducting qubits in which 
the freedom of superconducting phase is utilized as a bit (flux qubit) 
\cite{Mooij,Orlando,Wal,Chiorescu1,Majer,Izmalkov,Chiorescu2,
Yamashita,Ioffe,Blatter,Makhlin}.  
Three-junction type of the flux qubit proposed by Mooij {\it et al.} consists of 
a superconducting loop with three Josephson junctions, and the bonding and antibonding states 
formed by applying an external magnetic field are used as a bit 
\cite{Mooij,Orlando,Wal,Chiorescu1,Majer,Izmalkov,Chiorescu2}.  
For this qubit, single-qubit rotation (Rabi oscillation), 
the direct coupling between the two qubits, and the entangled states 
have been demonstrated \cite{Chiorescu1,Majer,Izmalkov,Chiorescu2}.  
A qubit with zero and ferromagnetic $\pi$ junctions, which requires no external magnetic field 
for the formation of the coherent two states, has been proposed \cite{Yamashita}.  
As another flux qubit, 
there are ``quiet" qubits consisting of $s$-wave/$d$-wave superconducting junctions or 
five Josephson junctions including one ferromagnetic $\pi$ junction \cite{Ioffe,Blatter}.  
In the quiet qubit, no current flows (quiet) in the system during the operation and 
therefore it is expected to be robust to the decoherence by the environment, whereas 
the difficulty of fabrication has been pointed out \cite{Makhlin}.

Furthermore, recent advances in the microfabricating techniques have promoted 
extensive works on spin-electronics \cite{Maekawa1,Maekawa2}.  
In particular, ferromagnetic $\pi$ junctions have been studied actively 
as a superconducting spin-electronic device.  
\cite{Buzdin,Radovic2,Demler,Ryazanov1,Kontos2,Sellier,Ryazanov2,Ryazanov3,Kontos3,Bauer}.  
In a superconductor/ferromagnet (SC/FM) junction, 
the superconducting order parameter penetrates into FM and 
oscillates due to the exchange field in FM \cite{Demler}.  
The system is stable at the phase difference equal to $\pi$ 
when the order parameters in two SC's take different sign in a SC/FM/SC junction, 
and this state is called ``$\pi$ state".  
The $\pi$ state is also explained 
from the point of view of spin-split Andreev bound states \cite{Sellier}.  
There are several experimental observations of the $\pi$ state 
in junction-type and SQUID-type structures 
\cite{Ryazanov1,Kontos2,Sellier,Ryazanov2,Ryazanov3,Kontos3,Bauer}.  
In the junction-type geometries, 
cusp structures in the temperature and FM's thickness dependences of the critical current 
have been observed as an evidence for the transition between 
the $\pi$ state and ordinary zero state \cite{Ryazanov1,Kontos2,Sellier,Ryazanov2}.  
More recently, the $\pi$ state has been confirmed via the reversed current-phase relation 
\cite{Ryazanov2}.  
In SQUID-type structures consisting of a superconducting ring with zero and $\pi$ junctions, 
it has been reported that 
the magnetic field dependence of the critical current is $\pi$-shifted 
compared to that in the ring with two zero junctions \cite{Ryazanov3,Kontos3}.  
In a superconducting ring with a single $\pi$ junction, an asymmetry 
in a magnetic field dependence of a spontaneous current 
due to the $\pi$ state has been observed \cite{Bauer}.

In this Letter, we propose a new qubit with 
two ordinary zero junctions and one ferromagnetic $\pi$ junction.  
We show that the qubit does not require an external magnetic field 
for forming the coherent two states, and only small external field is needed for 
distinguishing the states.  
This feature makes it possible to construct the qubit with smaller size, 
and therefore this qubit is advantageous to large-scale integration and 
expected to have a long decoherence time.

\begin{figure}[t]
\includegraphics[width=0.4\columnwidth]{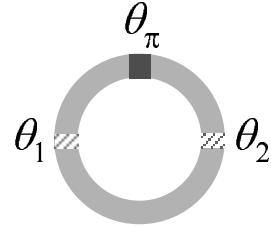}
\caption{Schematic diagram of a $\pi$ qubit consisting of 
a superconducting ring with two ordinary zero junctions and a $\pi$ junction.  
The phase differences at the zero junctions and the $\pi$ junction are 
$\theta_{1(2)}$ and $\theta_{\pi}$, respectively.}
\label{Fig1}
\end{figure}
We consider a superconducting ring with three Josephson junctions as shown in Fig. \ref{Fig1}.  
The two Josephson junctions are ordinary zero junctions 
with the phase differences $\theta_1$, $\theta_2$, and 
the other is a ferromagnetic $\pi$ junction with the phase difference $\theta_{\pi}$.  
The Hamiltonian in this system is expressed as $H = T + U$, 
where $T$ is an electrostatic energy term written as 
\begin{eqnarray}
T &=& - \frac{{4E_{C0} }}{{E_{C\pi } + 2E_{C0} }} \nonumber\\
&\times& \!\!\!\! \left[ {\left( {E_{C\pi } + E_{C0} } \right)
\left( {\frac{{\partial ^2 }}{{\partial \theta _1 ^2 }} 
+ \frac{{\partial ^2 }}{{\partial \theta _2 ^2 }}} \right) 
- 2E_{C0} \frac{{\partial ^2 }}{{\partial \theta _1 \theta _2 }}} \right], 
\end{eqnarray}
where $E_{C0(\pi)} = e^2/2C_{0(\pi)}$ is the Coulomb energy 
for a single charge at the zero ($\pi$) junction, 
$C_{0(\pi)}$ is the capacitance of the zero ($\pi$) junction 
and $e$ is the elementary charge.  
The term $U$ indicates a potential energy expressed as 
\begin{eqnarray}
U &=& - E_0 \left( {\cos \theta_1 + \cos \theta_2 }\right) 
- E_{\pi} \cos \left( {\theta_{\pi} + \pi} \right)
\nonumber\\
&+& \frac{{\left( {\Phi - \Phi_{ext} } \right)^2 }}{2L_s}, 
\label{U1}
\end{eqnarray}
where $E_0$ is the Josephson coupling constant in the zero junctions, 
$E_{\pi}$ is that in the $\pi$ junction, $\Phi$ is the total flux in the ring, 
$\Phi_{ext}$ is the external flux, and $L_s$ is the self inductance of the ring.  
Because of a single-valued wave function around the ring, 
the total flux and the phase differences satisfy the relation 
$\theta_1 + \theta_2 + \theta_{\pi} = 2\pi \Phi/\Phi_0 - 2\pi l$, 
where $\Phi_0$ is the unit flux and $l$ is an integer.  
Substituting the relation in Eq. (\ref{U1}), we obtain 
\begin{eqnarray}
U &=& - E_0 \left( {\cos \theta _1 + \cos \theta _2 }\right) \nonumber\\
&+& E_{\pi} \cos \left( {2\pi \frac{\Phi }{{\Phi _0 }} - \theta _1  - \theta _2 } \right) 
+ \frac{{\left( {\Phi  - \Phi _{ext} } \right)^2 }}{{2L_s }}.  
\label{U2}
\end{eqnarray}
From the condition that $U$ is minimum with respect to $\Phi$, 
i.e., $\partial U/\partial\Phi = 0$, 
we obtain the self-consistent equation as follows: 
\begin{eqnarray}
\alpha\beta \sin \left( {2\pi \frac{\Phi }{{\Phi _0 }} - \theta_1 - \theta_1 } \right) 
= 2\pi \left( {\frac{\Phi }{{\Phi _0 }} - \frac{{\Phi _{ext} }}{{\Phi _0 }}} \right), 
\label{sceq}
\end{eqnarray}
where $\alpha = E_{\pi}/E_0$ and $\beta = {4\pi ^2 E_0  L_s }/{\Phi _0 ^2 }$.  
By solving Eq. (\ref{sceq}) numerically, 
we obtain $\Phi = \Phi (\theta_1,\theta_2)$ as a function of $\theta_1$ and $\theta_2$.  
Substituting the obtained $\Phi (\theta_1,\theta_2)$ in the expression for $U$ (Eq. (\ref{U2})), 
we get $U = U (\theta_1,\theta_2)$ as a function of $\theta_1$ and $\theta_2$.  
Throughout this Letter, we assume $E_{C\pi} = E_{C0}/\alpha$ and 
use $\alpha = 0.8$, $\beta = 3.0 \times 10^{-3}$.  
The value of $\alpha$ is controllable by 
changing the contact area and the thickness of the insulators and of the ferromagnet, and 
that of $\beta$ is reasonable for the micrometer-size ring and 
the Josephson junction with several hundred nanoampere of the critical current.

\begin{figure}
\includegraphics[width=0.9\columnwidth]{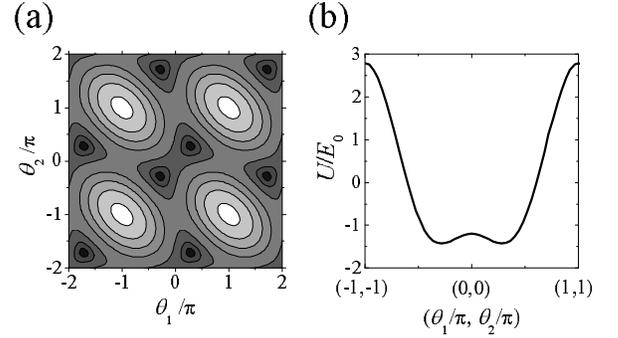}
\caption{(a) Contour plot of $U/E_0$ in the phase space 
without an external magnetic flux ($\Phi_{ext} = 0$).  
The lighter and darker parts correspond to 
the larger and smaller values of $U/E_0$, respectively.  
(b) The phase dependence of $U/E_0$ 
in the diagonal direction from $(\theta_1/\pi,\theta_2/\pi) = (-1,-1)$ to $(1,1)$.}
\label{Fig2}
\end{figure}
Figures \ref{Fig2}(a) and (b) show the $\theta_1$, $\theta_2$ dependence 
of $U$ without an external magnetic flux ($\Phi_{ext} = 0$).  
As shown in these figures, $U$ has degenerate two states 
$\left|\uparrow\right\rangle$ 
at near $(\theta_1/\pi,\theta_2/\pi) = (-1/2 + 2m,-1/2 + 2n)$ 
and $\left|\downarrow\right\rangle$ at near $(1/2 + 2m,1/2 + 2n)$ 
in the phase space where $m$ and $n$ are integers.  
At the $\left|\uparrow\right\rangle$ and $\left|\downarrow\right\rangle$ states, 
the currents of magnitude $\approx 0.8 I_0$ with clockwise and anticlockwise direction 
flow in the ring, respectively, where $I_0$ is the critical current in the zero junctions.  
Because of quantum tunneling between the degenerate 
$\left|\uparrow\right\rangle$ and $\left|\downarrow\right\rangle$ states, bonding 
$\left| 0 \right\rangle \propto \left|\uparrow\right\rangle + \left|\downarrow\right\rangle$ 
and antibonding 
$\left| 1 \right\rangle \propto \left|\uparrow\right\rangle - \left|\downarrow\right\rangle$ 
states which are used as a quantum bit are formed.  
Both at the $\left| 0 \right\rangle$ and $\left| 1 \right\rangle$ states, no current flows 
because the $\left|\uparrow\right\rangle$ and $\left|\downarrow\right\rangle$ components 
with the equivalent weight exist in the both states.  
The energy gap $\Delta E$ 
between the $\left| 0 \right\rangle$ and $\left| 1 \right\rangle$ states appears due to 
the quantum tunneling, and 
the existence of these states is confirmed by the microwave resonance 
with the frequency $\nu = \Delta E/h$, where $h$ is the Plank constant.  
From the numerical calculation for $E_0/E_{C0} = 30$, 
$\Delta E \approx 2.9 \times 10^{-24} \, {\rm J}$, which 
corresponds to the frequency $\nu \approx 4.4 \, {\rm GHz}$.  
This value of $\nu$ is comparable with that for the three-junction qubit 
\cite{Mooij,Orlando,Wal,Chiorescu1,Majer,Izmalkov,Chiorescu2}.  
\begin{figure}
\includegraphics[width=0.9\columnwidth]{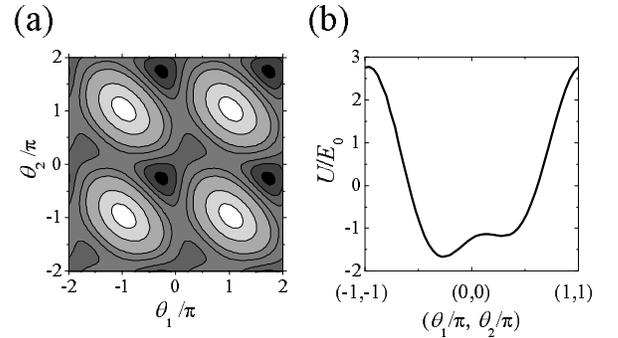}
\caption{(a) Contour plot of $U/E_0$ in the phase space 
with an external magnetic flux $\Phi_{ext} = 0.05 \Phi_0$.  
The lighter and darker parts correspond to 
the larger and smaller values of $U/E_0$, respectively.  
(b) The phase dependence of $U/E_0$ in the diagonal direction
from $(\theta_1/\pi,\theta_2/\pi) = (-1,-1)$ to $(1,1)$.}
\label{Fig3}
\end{figure}
Figures \ref{Fig3}(a) and (b) show the $\theta_1$, $\theta_2$ dependence 
of $U$ within an external magnetic flux $\Phi_{ext} = 0.05 \Phi_0$.  
As shown in these figures, 
the degeneracy of the $\left|\uparrow\right\rangle$ and $\left|\downarrow\right\rangle$ states 
is lifted due to the finite external magnetic flux.  
In this case, the $\left|\uparrow\right\rangle$ component increases and the 
$\left|\downarrow\right\rangle$ component decreases in the bonding $\left|0\right\rangle$ state, 
and vice versa in the antibonding $\left|1\right\rangle$ state.  
Therefore, finite spontaneous currents with the clockwise and anticlockwise directions 
flow at the $\left|0\right\rangle$ and $\left|1\right\rangle$ states, respectively, and 
the corresponding magnetic fluxes are induced in the ring.  
One can distinguish the states of the qubit through the measurements of the spontaneous flux 
by a SQUID located around the ring.  
As shown above, the qubit incorporating the $\pi$ junction 
does not require an external magnetic field for the formation of the coherent two states, 
and require only a small field around zero for detecting the states.  
A required field for the manipulation is of the order of a millitesla even if 
the dimension of the qubit is several 100 nm's.  
By this feature, (i) the qubit is advantageous to a large-scale integration, and 
(ii) it becomes easier to construct a smaller size of the qubit, which is robust to 
the decoherence effect due to the coupling to the environments.  
The $\pi$ qubit with the two Josephson junctions proposed in ref.\cite{Yamashita} requires 
a metallic-contact $\pi$ junction, whereas the present qubit works when 
the interface of the $\pi$ junction is insulating as well as metallic 
due to the two zero junctions.

For the quantum computation, a construction of an universal gate is needed.  
Most popular configuration of the universal gate consists of 
single-qubit rotation and controlled-NOT (CNOT) gates.  
The single-qubit rotation gate is done by 
applying a microwave with the frequency $\nu = \Delta E/h$ (Rabi oscillation).  
The CNOT gate is also realized as follows.  
Here we consider ``qubit A" as a control bit and ``qubit B" as a target bit 
under a small external magnetic field.  
Because of the magnetic interaction between the spontaneous fluxes in the two qubits, 
the energy gap in qubit B depends on the state of qubit A 
and is expressed as $\Delta E_{\rm{B0(B1)}}$ 
when qubit A is in the $\left|0\right\rangle (\left|1\right\rangle)$ state.  
By applying the microwave with the frequency $\Delta E_{\rm B1}/h$ to qubit B, 
the state of qubit B changes through the Rabi oscillation 
only when qubit A is in the $\left|1\right\rangle$ state.

In summary, we have proposed a new qubit consisting of a superconducting ring with 
two zero junctions and a single $\pi$ junction.  
In the system, the potential energy has double minima in the phase space without 
external magnetic fields because of the competition between the zero and $\pi$ states.  
The bonding and antibonding states (coherent states) are formed due to 
the quantum tunneling between the two degenerate states, 
and the coherent states are used as a bit in the qubit.  
A small external magnetic field around zero is needed for manipulating the state of the qubit.  
These features lead to a large-scale integration and a smaller size of the qubit 
which is resistant to the decoherence by the external noise and has a long decoherence time.

T.Y. was supported by JSPS Research Fellowships for Young Scientists.  
This work was supported by NAREGI Nanoscience Project, Ministry of 
Education, Culture, Sports, Science and Technology (MEXT) of Japan, 
and by a Grant-in-Aid from MEXT and NEDO of Japan.

\end{document}